\documentclass[aps,pra,twocolumn,showpacs,letterpaper,superscriptaddress]{revtex4-1}

\usepackage[colorlinks=true, citecolor=blue, linkcolor=blue, urlcolor=blue]{hyperref}
\usepackage{graphicx,dcolumn,longtable,epsfig}
\usepackage[usenames]{color}
\usepackage{amssymb}
\usepackage{amsmath}
\usepackage{bm}
\usepackage{footnote}
\usepackage{float}
\usepackage{subfigure}
\usepackage{color}

\usepackage{ulem}
\usepackage[T1]{fontenc}

\newcommand{\be}{\begin{equation}}
\newcommand{\ee}{\end{equation}}

\def\bea{\begin{eqnarray}}
\def\eea{\end{eqnarray}}

\begin{document}
\title{Quantum phases of lattice dipolar bosons coupled to a high-finesse cavity}
\author{Yaghmorassene Hebib }
\email{These authors contributed equally: Y. Hebib, C. Zhang}
\affiliation{Department of Physics, Clark University, Worcester, Massachusetts 01610, USA}
\author{Chao Zhang}
\email{These authors contributed equally: C. Zhang, Y. Hebib}
\email{zhangchao1986sdu@gmail.com}
\affiliation{Department of Modern Physics, University of Science and Technology of China, Hefei, Anhui 230026, China}
\author{Jin Yang}
\email{dypole_jin@mit.edu}
\affiliation{Department of Physics, Research Laboratory of Electronics, MIT-Harvard Center for Ultracold Atoms, Massachusetts Institute of Technology, Cambridge, Massachusetts 02139, USA}
\author{Barbara Capogrosso-Sansone}
\affiliation{Department of Physics, Clark University, Worcester, Massachusetts 01610, USA}

\begin{abstract}
Two types of long range interactions, dipolar interaction and cavity-mediated interaction lead to exotic quantum phases. Both interactions have been realized and observed in optical lattice setups. Here, we study quantum phases of dipolar bosons trapped in optical lattices and coupled to a high-finesse cavity where both dipolar interaction and cavity-mediated interaction coexist. We perform quantum Monte Carlo simulations, and find that the checkerboard solid is enhanced and the checkerboard supersolid phase can exist in a wide range of densities (e.g. $ 0.27\lesssim n\lesssim0.73 $). Our unbiased numerical results suggest that both solid and supersolid phases can be achieved experimentally with magnetic atoms coupled to a cavity.

\end{abstract}

\maketitle

%%%%%%%%%%%%%%%%%%%%%%%%%%%%%%%%%%%%%%%%%%%%%%%%%%%%%%%%%%%%%%%%%%%%%%%%%
\section{Introduction}

Competition between different interactions or energy scales is key to understand the  physics world. A secondary interaction is treated as a perturbation when it is small compared with the primary interaction. But when it changes to be dominant, the original eigenenergy and eigenwavefunction of the system need to reform correspondingly \cite{sakurai_modern_2017-3}. Quantum phase transitions also result from competition. In optical lattice experiments, at low temperatures, a transition from superfluid to Mott insulator can be realized when on-site interaction starts  dominating over tunneling~\cite{PhysRevB.40.546,Greiner_MI-SF,PhysRevLett.81.3108,PhysRevB.75.134302}. The competition between short-ranged, on-site interaction and long-range interaction gives rise to novel exotic quantum phases, like supersolid states \cite{EsslingerNature}.
%THE FIRST PARAGRAPH SHOULD BE EXPANDED WITH FOCUS ON EXPERIMENTS AND MORE REFERENCES SHOULD BE ADDED.

In optical lattices, two types of long-range interactions have been realized and observed - dipolar interaction and cavity-mediated long-range interaction~\cite{Baier2016,polarMol,RevModPhys.82.2313,Low_2012,EsslingerNature,Farokh:2021}. Theoretically, both types of long range interactions have been studied comprehensively, and numerous results on exotic quantum phases realized by these interactions have been reported~\cite{Danshita:2009cp,PhysRevLett.104.125301,PhysRevA.83.013627,CapogrossoSansone:2011eq,Bandyopadhyay:2019ew,Zhang:2018ew,Kraus:2020es,Zhang_2015,Giovannastagger,PhysRevA.103.043333, Wu:2020kf, PhysRevA.90.043604,PhysRevLett.84.1599,PhysRevA.90.043635,PhysRevA.105.063302,Li:2013bn,Niederle:2016fi,Dogra:2016hy,Chen:2016kv,Sundar:2016ie,PhysRevB.95.144501,Bogner:2019ij,Blass:2018iw,Zhang2020EPJB,Zhangfphy, carl2022, PhysRevA.106.063313,massimo2020}. So far, these two types of long-range interactions have been studied singly so that a study on quantum phases in a system with both dipolar and cavity-mediated long range interactions is still absent today. Experimentally, creating optical lattices and conducting experiments within cavities is not challenging and has recently attracted lot of attention \cite{ritsch_cold_2013, aspelmeyer_cavity_2014, muniz_exploring_2020}.

In this manuscript, we are interested in a gas of dipolar bosons trapped in an optical lattice and coupled to a high-finesse cavity. The particles in the system interact via dipolar interaction and infinite-range, cavity-mediated interaction.  %From a theoretical standpoint, lattice bosons with 
Dipolar interactions are known to stabilize a plethora of charge density waves and supersolid phases, and cavity-mediated interactions have been shown to stabilize charge density waves between odd and even sites. Here, we study the ground-state phase diagram of lattice dipolar bosons in the presence of global-range interactions by means of quantum Monte Carlo simulations. On the one hand, we are interested in understanding  how the presence of photon-mediated interactions affects the quantum phases stabilized in dipolar bosons trapped in optical lattices; on the other hand, we want to quantify the changes in the phase diagram of lattice bosons with cavity-mediated interactions when dipolar interaction is switched on. For the range of parameter studied, we find that the presence of cavity-mediated infinite-range  interactions enhances robustness of the checkerboard solid and supersolid phases. Interestingly, the checkerboard supersolid can survive to filling factors as low as $\sim 0.27$, which is comparable to what currently achievable experimentally with polar molecules~\cite{christakis_probing_2022}. Moreover, cavity-mediated interactions significantly lower the dipolar interaction strength needed to observe checkerboard supersolid, facilitating the observation of such phase with magnetic atoms.

This paper is organized as follows: In Sec.~\ref{sec:sec2} we introduce the Hamiltonian of the system. In Sec.~\ref{sec:sec3} we discuss various phases and the corresponding order parameters. In Sec.~\ref{sec:sec5} and~\ref{sec:sec4} we present the phase diagrams of the above system in the soft core and hard core case separately. 
In Sec.~\ref{sec:sec6} we discuss the experimental realization. We conclude the paper in Sec.~\ref{sec:sec7}.

\section{Hamiltonian}
\label{sec:sec2}

\begin{figure}[h]
\includegraphics[width=0.45\textwidth]{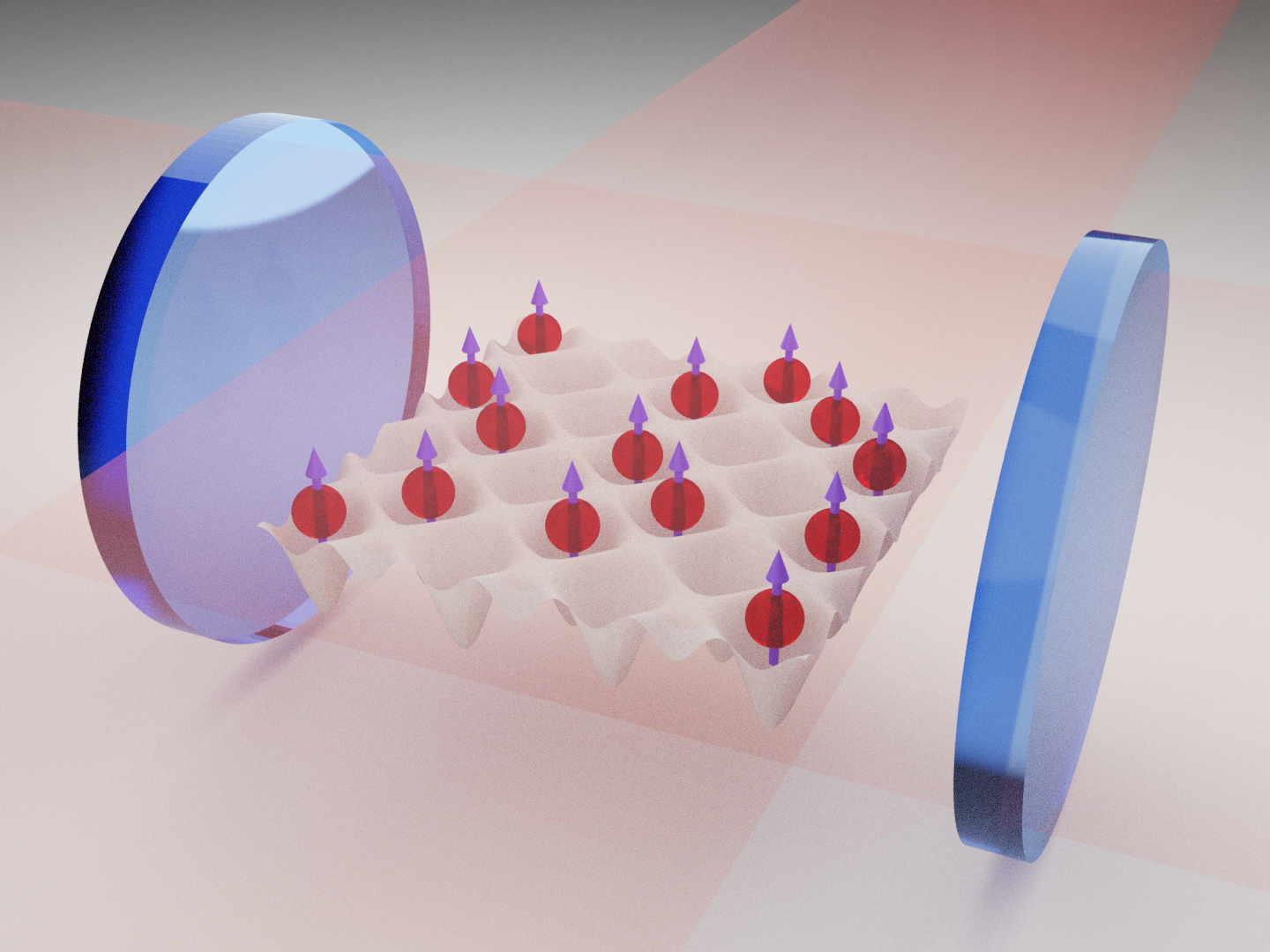}
\caption{Schematic representation of the system. Dipoles are trapped in a two-dimensional optical lattice and are aligned parallel to each other along the direction of polarization, determined by an electric/magnetic field. The polarization is along the $z$-axis. The lattice is coupled to a high-finesse optical cavity which is represented by two mirrors. 
} 
\label{setup}
\end{figure}

We consider a gas of dipolar bosons trapped in a two-dimensional square optical lattice inside a high-finesse optical cavity (see Fig.~\ref{setup}). Dipoles are aligned perpendicular
to the optical lattice plane so that the dipolar interaction is purely repulsive and isotropic. In the single band approximation, the system can be described by the extended Bose-Hubbard model~\cite{EsslingerNature,Dogra:2016hy}:

%\begin{equation}
\begin{align}
\nonumber H& =-J\sum_{\langle i, j\rangle } a_i^\dagger a_j + \frac{U_\text{s}}{2}\sum_i n_i(n_i-1) + \frac{V_{\text{dip}}}{2} \sum_{ i,j } \frac{n_i n_j}{r_{ij}^3} \\
&-\frac{V_\text{ca}}{L^2} \bigg{(} \sum_{i \in e} n_i- \sum_{j \in o} n_j  \bigg{)}^2 - \mu \sum_i n_i \;\; ,
\label{Eq1}
\end{align}
%\end{equation}
where the first term is the kinetic energy characterized by the hopping amplitude $J$. Here $\langle \cdot \rangle$ denotes nearest neighboring sites, $a^{\dagger} (a)$
bosonic creation (annihilation) operators satisfying the bosonic commutation relations. The second term is the short-range on-site repulsive interaction with interaction
strength $U_s$. Here, $n_i= a^{\dagger}_{i}a_{i}$ is the particle number operator. The third term is the dipolar interaction term and $r_{ij}=|\bf r_i-\bf r_j|$ is the relative distance between site $i$ and site $j$.  
The fourth term is the cavity-mediated long-range interaction with interaction strength $V_\text{ca}$. The summations $i \in e$ and $j\in o$ denote summing over even and odd lattice sites respectively, where even (odd) sites refer to even (odd) values of the sum of $x$ and $y$ lattice site coordinates. This term favors population imbalance between even and odd sites. A derivation of this term can be found in~\cite{EsslingerNature}. Finally, $\mu$ is the chemical potential.

In the following, we present unbiased results of phase diagrams of Hamiltonian~(\ref{Eq1}) based on path-integral quantum Monte Carlo using the worm algorithm~\cite{PROKOFEV1998253}. We have performed the simulations on an $L\times L=N_{\rm s}$ square lattice system with $L=16,20,24,30$ (we choose the lattice constant $a$ to be our unit of length). We have imposed periodic boundary conditions in both spatial dimensions. We use Ewald summation to account for the long-range dipolar interaction.  The inverse temperature $\beta$ is set to $\beta=L$.

\section{Order parameters}
\label{sec:sec3}

In this section, we describe the order parameters used to characterize superfluid (SF) phase, checkerboard solid (CB) phase, checkerboard supersolid (CBSS) phase. Specifically, we calculate superfluid density $\rho_s$ and structure factor $S(\pi, \pi)$. The superfluid density is calculated in terms of the winding number~\cite{Ceperley:1989hb}: $\rho_s=\langle \mathbf{W}^2 \rangle /DL^{D-2}\beta$, where $\langle\mathbf{W}^2\rangle=\sum_{i=1}^D\langle W_i^2\rangle$ is the expectation value of winding number square, $D$ is the dimension of the system and here $D=2$, $L$ is the linear system size, and $\beta$ is the inverse temperature. The structure factor characterizes diagonal long-range order and is defined as: $S(\mathbf{k})=\sum_{\mathbf{r},\mathbf{r'}} \exp{[i \mathbf{k}\cdot (\mathbf{r}-\mathbf{r'})]\langle n_{\mathbf{r}}n_{\mathbf{r'}}\rangle}/N$, where $N$ is the particle number. $\mathbf{k}$ is the reciprocal lattice vector. We use $\mathbf{k}=(\pi, \pi)$ to identify a checkerboard density pattern. Notice that, in the CBSS, $\rho_s$ and $S(\pi,\pi)$ are finite simultaneously.

Another quantity we monitor is compressibility  defined as $\frac{\beta\Delta N^2}{L^2}$, where $\Delta N^2=\langle(N-\langle N\rangle)^2\rangle$. The compressibility is finite for compressible phases and zero (in the thermodynamic limit) for incompressible phases. As we shall discuss below, for large enough interactions, we find a variety of incompressible phases.

\section{Softcore case}
\label{sec:sec5}

\begin{figure}[t]
\includegraphics[width=0.45\textwidth]{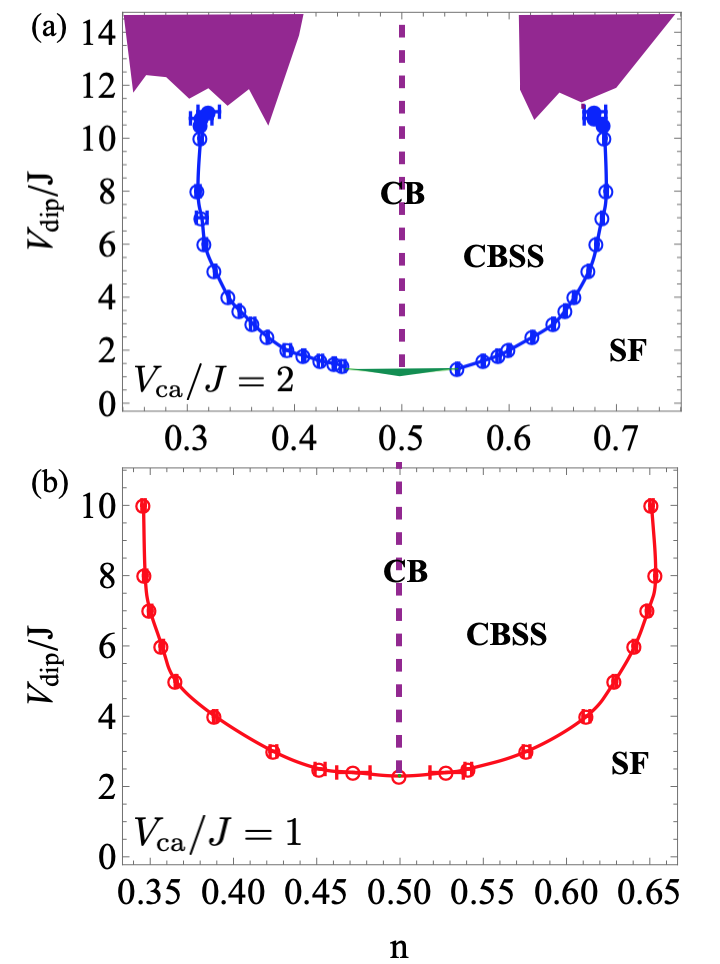}
\caption{ Softcore case: Phase diagram of Hamiltonian~(\ref{Eq1}) as a function of $V_{\rm dip}/J$ and particle density $n$, computed via quantum Monte Carlo simulations at $U_{\rm s}/J = 20$. (a): $V_{\rm ca}=2J$; (b): $V_{\rm ca}=J$.   We observe a checkerboard (CB) solid at $n=0.5$ (dashed purple line), a checkerboard supersolid (CBSS), a superflluid (SF) phase, and an incompressible phase (IP, purple region). The  green solid region at the tip of the lobe in (a) marks the first-order phase transition. Blue and red open circles are second-order transition points calculated using finite size scaling (see text for details). Filled circles represent first-order phase transition points. 
} 
\label{PDsoft}
\end{figure}

\begin{figure}[t]
\includegraphics[width=0.45\textwidth]{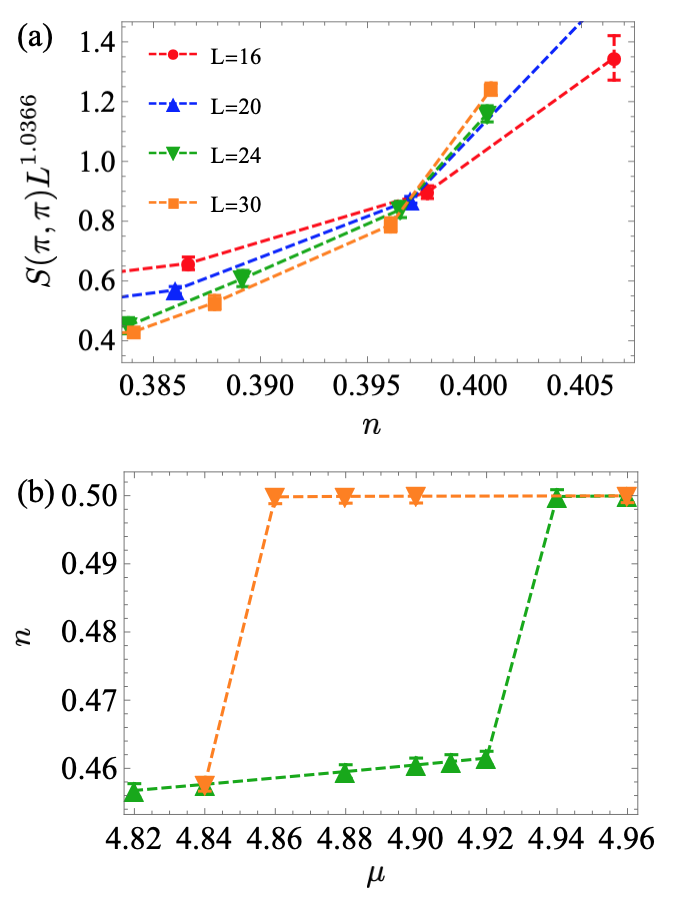}
\caption{Softcore case: (a) Finite size scaling of the structure factor $S(\pi, \pi) L^{1.0366}$ as a function of filling factor $n$ for system sizes $L=16$ (red circles), 20 (blue up triangles), 24 (green down triangles), and 30 (orange squares) at $V_{\text{dip}}=2J$ and $V_{\text{ca}}=2J$. The CBSS to SF phase transition happens at $n=0.397 \pm 0.03$. (b) The hysteresis curve of filling factor $n$ vs chemical potential $\mu$ at $V_{\rm dip}=1.3J$ and $L=20$. 
} 
\label{hyst}
\end{figure}

\begin{figure}[t]
\includegraphics[width=0.48\textwidth]{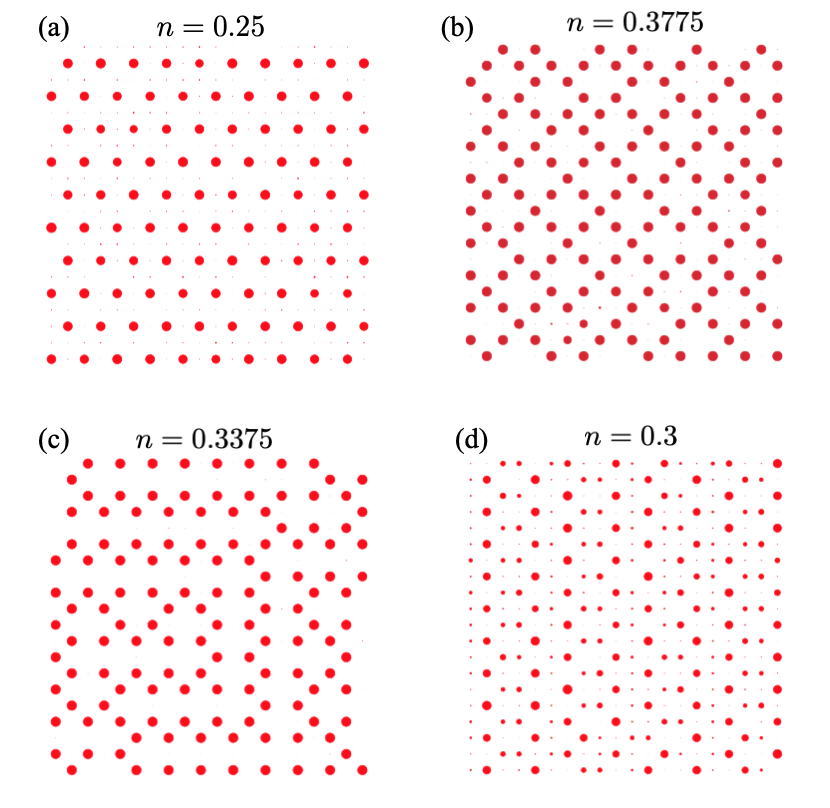}
\caption{Density maps for different fillings. Each circle corresponds to a single lattice site, and its radius is proportional to the local density. (a) $n=0.25$ (softcore and hardcore), (b) $n=0.3775$ (softcore), (c) $n=0.3375$ (hardcore), (d) $n=0.3$ (hardcore). 
}
\label{densitymap}
\end{figure}

\begin{figure}[t]
\includegraphics[width=0.48\textwidth]{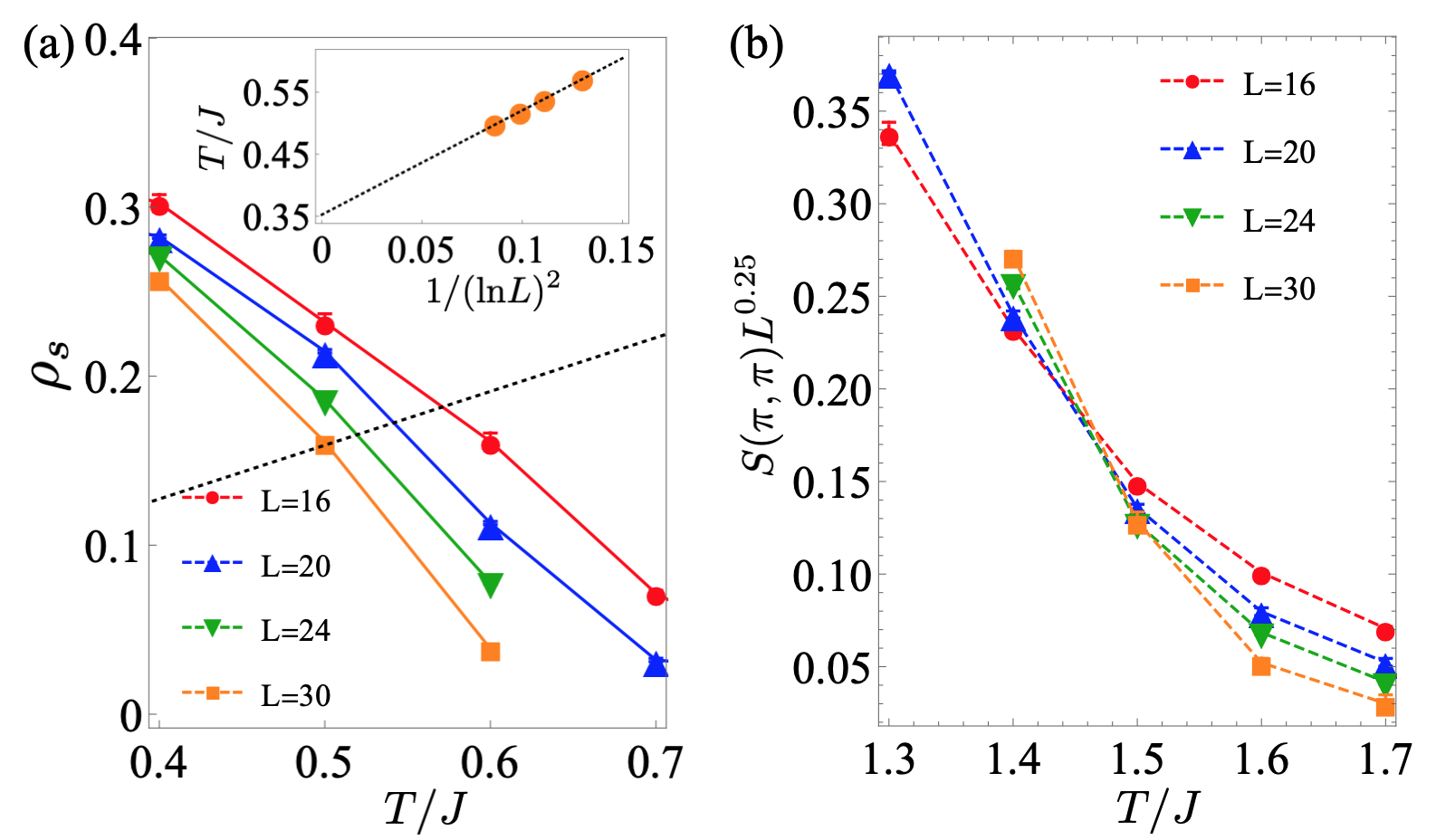}
\caption{Softcore case: $V_{\text{dip}}=1.8J$, $V_{\text{ca}}=2J$, and $n=0.5605$. Upon increasing the temperature, thermal fluctuations destroy the checkerboard supersolid phase in favor of a normal fluid in two steps. First, superfluidity is destroyed and the checkerboard supersolid becomes a checkerboard solid via a Kosterlitz–Thouless phase transition. Then, the checkerboard solid phase melts into a normal fluid via a two-dimensional Ising transition. In (a) we show $\rho_s$ as a function of $T/J$ for $L=16$ (red circles), 20 (blue up triangles), 24 (green down triangles), and 30 (orange squares). The dashed line is $T/J\pi$. Inset: intersection points between the $T/J\pi$ line and the $\rho_s$ versus $T/J$ curves for each $L$ are used to extract $T_c/J\sim0.35 \pm 0.01$. (b) scaled structure factor with $2\beta/\nu=0.25$ for $L=16$ (red circles), 20 (blue up triangles), 24 (green down triangles), and 30 (orange squares). The crossing determines the critical temperature $T_c/J = 1.47\pm0.03$. } 
\label{FiniteT}
\end{figure}

We present our results for the soft core case with fixed $U_{\rm s}/J=20$ and filling $n\le 1$. In Fig.~\ref{PDsoft}, we study how the presence of cavity-mediated interactions affects the phase diagram of a soft core dipolar system. In the absence of cavity interactions, the phase diagram in the $V_{\rm dip}$ vs $n$ plane features three solid phases corresponding to filling factors $n=1/2$, $3/4$, $1$ separated by a supersolid phase, a succession of incompressible states in the lower density regime stabilized at rational filling factors, a Mott insulator phase at unit filling and low dipolar interaction, and a SF phase~\cite{PhysRevA.90.043635}. In Fig.~\ref{PDsoft}, we show the phase diagram in the $V_{\rm dip}$ vs $n$ plane for cavity interaction $V_{\rm ca}/J=2$ (a) and $V_{\rm ca}/J=1$ (b). As expected, the CB solid (vertical purple dashed line) stabilized at half filling appears for lower values of the dipolar interaction as a finite cavity-mediated interaction favors the stabilization of a density-wave between even and odd sites. 
Specifically, CB is stabilized for $V_{\rm dip}\sim 1.13J$ at $V_{\rm ca}=2J$, and $V_{\rm dip}\sim 2.3J$ at $V_{\rm ca}=J$ compared to $V_{\rm dip}\sim 4.75J$ at $V_{\rm ca}=0$. 
%This is expected, considering that cavity-mediated interactions favor the stabilization of a density-wave between even and add sites. 

Upon doping the CB solid with particles or holes, the system enters a CBSS phase. For $V_{\rm ca}=0$, on the particle side, a supersolid exists in the full range $1/2<n<1$ (with the exception of $n=3/4$) with the density ordering which differs depending on how close the value of density is to the solids at $n = 1/2$, $3/4$, or $1$. Here, only CBSS is observed which disappears to a SF phase for large enough particle doping.  Similarly, the CBSS also appears for $n<1/2 $ and is destroyed for large enough doping. We notice that for lower values of dipolar interaction, the CBSS replaces the  succession of solids stabilized at rational filling factors observed in~\cite{PhysRevA.90.043635}. In other words, the CBSS on the hole side is more robust against doping compared to what observed in~\cite{PhysRevA.90.043635} and it can survive to filling factor as low as $n\sim 0.31$. 
The translational order of the CBSS is destroyed via a second-order phase transition belonging to the $(2+1)$ Ising universality class, leaving the system in an SF phase. The boundary between the SS and SF phases (red and blue open  circles in Fig.~\ref{PDsoft}(a) and (b)) is found using standard finite-size scaling. Specifically, we determine critical points using finite-size scaling for the static structure factor by plotting $S(\mathbf{\pi,\pi})L^{2\beta/\nu}$ versus density, with scaling coefficient ${2\beta/\nu}=1.0366$~\cite{PhysRevB.59.11471}. Critical points are determined from the intersection of $S(\mathbf{\pi,\pi})L^{2\beta/\nu}$ curves for different $L$’s (Fig.~\ref{hyst}(a)). Overall, we notice that the CBSS is very robust against doping as it can exist for a wide range of densities already at relatively small values of $V_{\rm dip}$. For example, excluding half-filling, CBSS exists within the ranges $0.375<n<0.62$ at $V_{\rm dip}=2.5J$ and $V_{\rm ca}=2J$, $0.31<n<0.69$ at $V_{\rm dip}=8J$ and $V_{\rm ca}=2J$, or $0.39<n<0.61$ $V_{\rm dip}=4J$ and $V_{\rm ca}=J$.

We are not able to resolve the transition between CB and SF at fixed half filling. At $V_{\rm ca}=2J$, we do observe a direct CB-SF first-order transition in the vicinity of the tip as confirmed by a discontinuity in density, structure factor and superfluid stiffness when crossing the CB-SF boundary at fixed $V_{\rm dip}$, and by the hysteretic behavior of the same quantities ((Fig.~\ref{hyst}(b))). The solid green area around the tip of lobe in Fig.~\ref{PDsoft}(a) correspond to the densities for which one would observe phase coexistence. It is well-known that first-order phase transitions are forbidden in dipolar systems, based on surface tension arguments. Rather, two distinct phases are separated by a macroemulsion phase~\cite{PhysRevB.70.155114}. Nonetheless, due to logarithmic size dependence in the surface tension, it is possible that, for all practical purposes, in a finite-size system the transition would effectively be of first order (see e.g. in~\cite{PhysRevA.103.043333,PhysRevA.105.063302}). 
For lattice bosons with onsite and infinite-range interactions and no dipolar interaction, first-order phase transitions have been observed before (see e.g~\cite{PhysRevB.95.144501,Zhangfphy}). Here, we also observe it in the presence of dipolar interaction.
 In Fig.~\ref{hyst}(b), we show the hysteresis curve of $n$ vs $\mu$ at $V_{\rm dip}=1.3J$ and $L=20$. We notice that hysteretic behavior is observed in a narrow range of $\mu$ of $\sim 2\%$.  

For $V_{\rm ca}=2J$, we checked under which conditions other incompressible phases appear in the phase diagram.  Considering the enhanced robustness of CBSS due to non-zero  $V_{\rm ca}$, in order to start observing solid phases  at rational fillings other than $n=1/2$ (purple solid region  in Fig.~\ref{PDsoft}(a)), we need  $V_{\rm dip}\sim 10.75J$ which is almost ten times larger than $V_{\rm dip}\sim1.13J$ corresponding to the onset of CB order. In comparison, for $V_{\rm ca}=0$, solids at other rational filling factors start appearing (on the hole side) at $V_{\rm dip}\sim 8J$~\cite{PhysRevA.90.043635}, approximately two times $V_{\rm dip}\sim 4.75J$ corresponding to the onset of CB order. As $V_{\rm dip}$ is increased, this succession of solids that we call incompressible phase (IP)  tends to become dense in the filling factor for a wide range of densities. In Fig.~\ref{densitymap}, we show examples of density maps within the purple region. Here, each circle corresponds to a single lattice site, and its radius is proportional to the local density. In Fig.~\ref{densitymap}(a), we show the star solid at $n=1/4$. For other densities, (see e.g. Fig.~\ref{densitymap}(b)), the density pattern possesses defects which appear in order to accommodate certain rational fillings. All these findings are very similar for $V_{\rm ca}=J$ (not shown here). It is interesting to notice that for dipolar strength corresponding to formation of solids at fillings other than $0.5$, the CBSS-SF transition appears to become of first order (filled circles in Fig.~\ref{PDsoft}(a)). Finally, we notice that,  at unit filling, the system is in a Mott insulator state for the range of dipolar interaction explored.

Let us now turn to a brief discussion on the robustness of the CBSS against thermal fluctuations. 
%We find the solid order to be the most robust against thermal fluctuations. 
Superfluidity  disappears via a Kosterlitz-Thouless transition~\cite{Kosterlitz:1973fc} while the solid order melts via a two-dimensional Ising transition. 
In Fig.~\ref{FiniteT}, we show the superfluid density $\rho_s$ as a function of $T/J$ for $L=16$, $20$, $24$, and $30$ (circles, up triangles, down triangles, squares respectively) at fixed  $n=0.5605$, $V_{\rm dip}=1.8J$, and $V_{\rm ca}=2J$. In the thermodynamic limit, a universal jump is observed at the critical temperature given by $\rho_s(T_c)=2m k_B T_c/\pi \hbar^{2}$.  Here, $m$ is the effective mass in the lattice, $m=\hbar^2/2Ja^2=1/2J$ in our units ($\hbar=1$, $k_B=1$, lattice step $a=1$). In a finite size system, this jump is smeared out as one can see in Fig.~\ref{FiniteT}(a). To extract the critical temperature in the thermodynamic limit, we apply finite-size scaling to $T_c(L)$. From renormalization-group analysis  one finds $T_c(L)=T_c(\infty)+\frac{c}{\ln^2(L)}$, where $c$ is a constant and $T_c(L)$ is determined from $\rho_s(T_c,L)=2m k_B T_c/\pi \hbar^{2}$~\cite{PhysRevLett.39.1201,Ceperley:1989hb}.
The dashed line in Fig.~\ref{FiniteT}(a) corresponds to $\rho_s=T/J\pi$  and its intersection points with each $\rho_s$ vs. $T/J$ curve are used to find $T_c$ as shown in the inset. We find $T_c/J=0.35 \pm 0.01$, a slightly higher temperature than $T_c/J \sim 0.25$ reported in~\cite{PhysRevA.105.063302}. 
Above this temperature the system is in a CB. The solid order melts in favor of a normal fluid via a two-dimensional Ising transition.  We use standard finite size scaling as shown in Fig.~\ref{FiniteT}(b), where we plot the scaled structure factor $S(\pi,\pi)L^{2\beta/\nu}$, with $2\beta/\nu=0.25$ as a function of $T/J$ for $L=16$, $20$, $24$, and $30$ (circles, up triangles, down triangles, squares respectively). The crossing indicates a critical temperature $T_c/J = 1.47\pm0.03$ making the CB solid considerably more robust against thermal fluctuations than what found in~\cite{PhysRevA.105.063302} where $T_c/J \sim 0.68$ has been reported.

\section{Hardcore case}
\label{sec:sec4}

\begin{figure*}[t]
\includegraphics[width=0.92\textwidth]{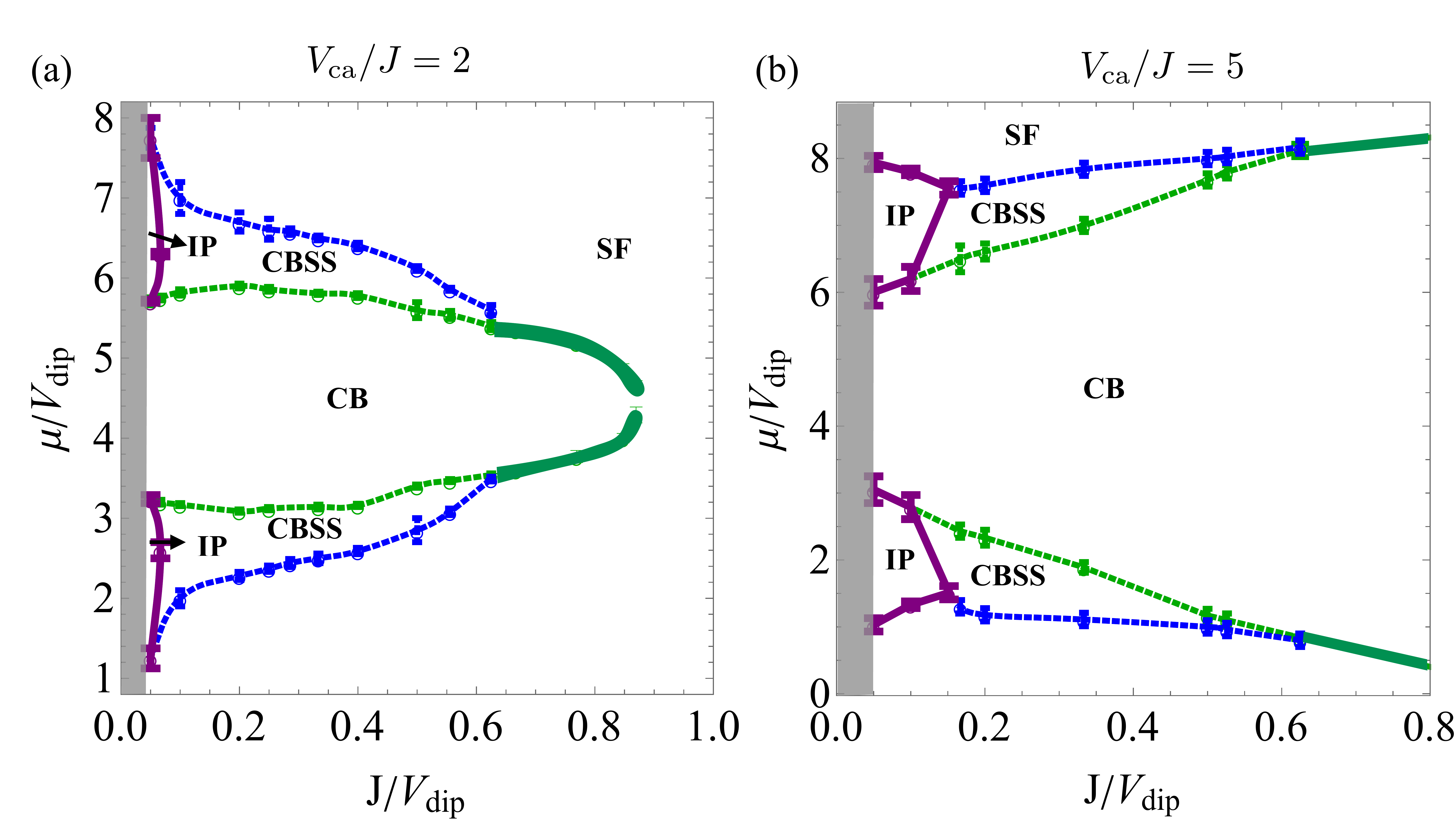}
\caption{Hardcore case: Ground state phase diagram of the system described by Eq.~\ref{Eq1} at fixed cavity interaction $V_{\text{ca}}/J=2$ (a) and $V_{\text{ca}}/J=5$ (b) in the plane $\mu/V_{\text{dip}}$ vs $J/V_{dip}$. We observe a superfluid phase (SF), checkerboard phase (CB), checkerboard supersolid phase (CBSS), and incompressible phase (IP). Dashed lines correspond to second-order phase transitions while solid-lines correspond to first-order phase transitions. The gray shadowed region corresponds to $J/V_{\text{dip}} \le 0.05$ which has not been explored. 
} 
\label{FIG1}
\end{figure*}

\begin{figure*}[t]
\includegraphics[width=0.92\textwidth]{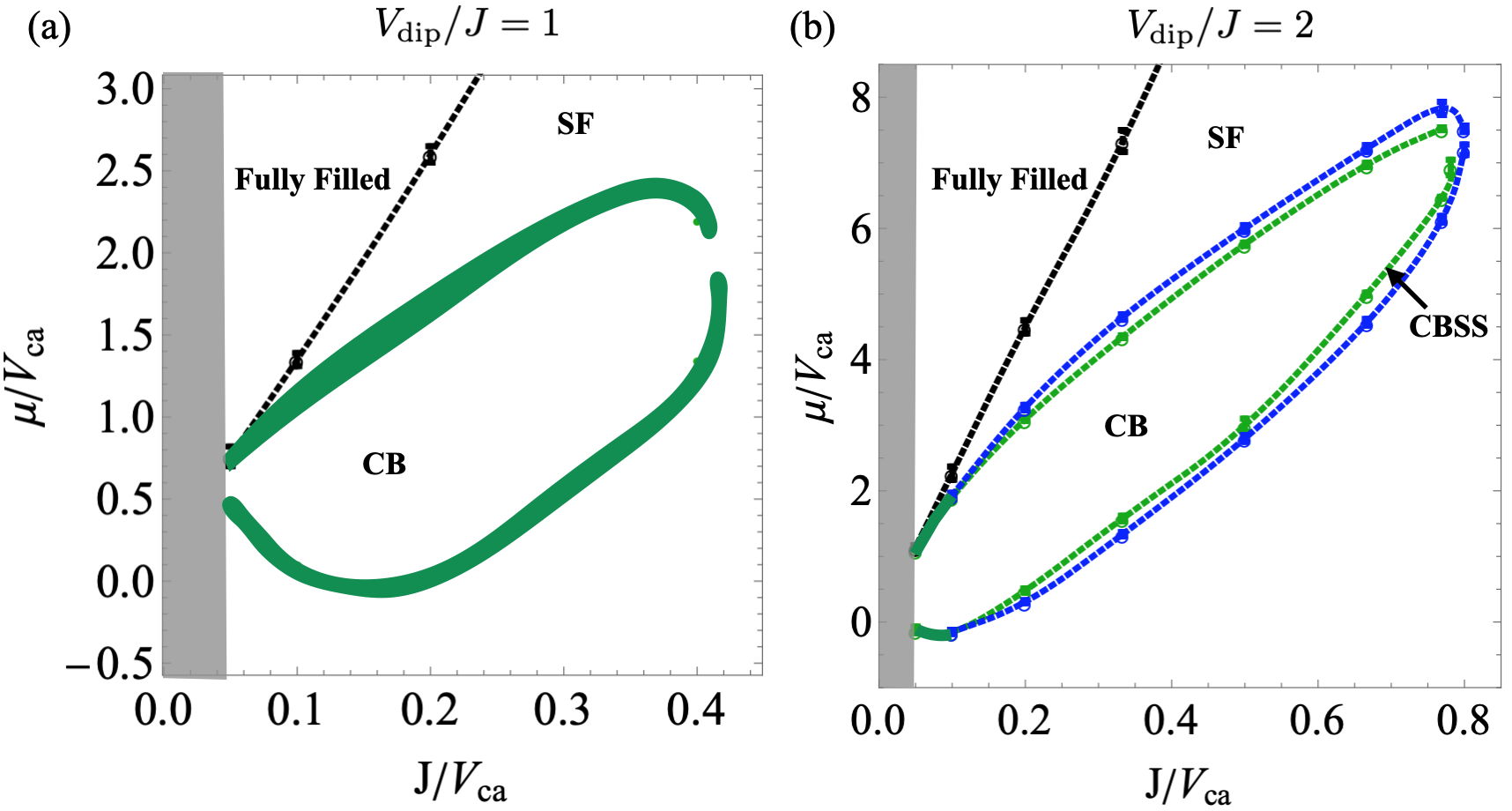}
\caption{Hardcore case: Ground state phase diagram of the system described by Eq.~\ref{Eq1} at fixed dipolar interaction $V_{\text{dip}}/J=1$ (a) and $V_{\text{dip}}/J=2$ (b) in the plane $\mu/V_{\text{ca}}$ vs $J/V_{\text{ca}}$. We observe a superfluid phase (SF), checkerboard solid phase (CB) and fully filled phase in (a); superfluid phase (SF), checkerboard phase (CB), checkerboard supersolid phase (CBSS), and fully filled phase in (b). Dashed lines correspond to second-order phase transitions while the solid thick line corresponds to first-order phase transition. The gray shadowed region corresponds to $J/V_{\text{ca}} \le 0.05$ which has not been explored. 
}
\label{FIG2}
\end{figure*}

In Fig.~\ref{FIG1}, we study how the presence of cavity-mediated interactions affects the phase diagram of a hardcore dipolar system. In the absence of cavity interactions, the phase diagram in the $\mu/V_{\rm dip}$ vs $J/V_{\rm dip}$ features three main lobes (we discuss densities $n\le0.5$, considering the particle-hole symmetry of the system) corresponding to solids at filling factors $n=1/2$, $1/3$, $1/4$, a SF phase, and a supersolid phase stabilized upon doping the solids with particles or holes~\cite{PhysRevLett.104.125301}. Moreover, for  $V_{\rm dip}\gtrsim50 J$ a succession of incompressible states at rational filling factors is stabilized. In Fig.~\ref{FIG1}, we show the phase diagram in the $\mu/V_{\rm dip}$ vs $J/V_{\rm dip}$ at $V_{\rm ca}=2J$ (a) and $V_{\rm ca}=5J$ (b). 

Let us start our discussion with Fig.~\ref{FIG1}(a), corresponding to $V_{\rm ca}=2J$. The infinite-ranged cavity-mediated interaction favors density-waves between even and odd sites. As a result, the lobe corresponding to the stripe solid at filling $1/3$ observed in~\cite{PhysRevLett.104.125301} has disappeared. 
The main lobe in Fig.~\ref{FIG1}(a) corresponds to the CB solid. The CB order appears at $ V_{\rm dip} \sim 1.15J$ compared to $ V_{\rm dip} \sim 3.6J$ at $V_{\rm ca}=0$. This is expected, considering that cavity-mediated interactions favor the stabilization of a CB solid. For $V_{\rm dip}\ge 1.6J$, the CB lobe is surrounded by a CBSS. This supersolid phase survives for densities $0.34<n<0.66$ ($n= 0.5$ excluded) at $V_{\rm dip}=4J$, and $0.4<n<0.6$  ($n=0.5$ excluded)  at $V_{\rm{dip}}=1.8J$ making the CBSS more robust against doping than in the case $V_{\rm ca}=0$.
A major difference with the phase diagram at $V_{\rm ca}=0$ is that there exists a range of dipolar interaction for which a CBSS does not intervene between the solid and the SF phase upon doping the solid with particles or holes.  Instead, we found evidence of a first-order phase transition (solid thick green line) as confirmed by a discontinuity in density, structure factor, and superfluid stiffness when crossing the CB-SF boundary at fixed $V_{\rm dip}$, and hysteretic behavior of the same quantities. We notice that when crossing the CB-SF boundary at fixed filling $n=1/2$, we were not able to resolve neither a  first-order transition nor a microemulsion phase.

For large enough $V_{\rm dip}$ ($V_{\rm dip}\gtrsim 15 J$) a variety of incompressible phases appears, the solid at $n=1/4,3/4$ being the first ones appearing. In Fig.~\ref{densitymap}(a) we show the density map at $n=1/4$ which corresponds to a star solid, likewise what is observed in the absence of cavity-mediated interactions. Other incompressible phases are stabilized at other rational filling factors. We find that in many cases (see e.g. Fig.~\ref{densitymap}(c)) the density pattern possesses defects (which appear in order to accommodate certain rational fillings). In other cases, we observe some of the particles being delocalized within a pair of sites but with no global phase coherence (Fig.~\ref{densitymap}(d)).

As the cavity interaction is increased to $V_{\rm ca}/J=5$ (Fig.~\ref{FIG1} (b)), only quantitative changes are observed: The CB solid at half filling is always stabilized for any  $V_{\rm dip}$; the CBSS disappears in favor of the IP at lower values of $V_{\rm dip}$; the CBSS is more robust against doping (e.g., at $V_{\rm dip}=5.0J$, CBSS survives for densities within the range $0.27\lesssim n \lesssim 0.73$).

In Fig.~\ref{FIG2},  we study how the presence of dipolar interactions affects the phase diagram of hardcore bosons in a cavity. In the absence of dipolar interactions, the phase diagram in the $\mu/V_{\rm ca}$ vs $J/V_{\rm ca}$~\cite{Zhangfphy} features a CB solid, a SF phase, and a coexistence region separating the CB from the fully filled lattice and the SF phase. The coexistence region results from robust first order phase transition between CB and SF or CB and fully filled lattice.     In Fig.~\ref{FIG2}, we show how this phase diagram changes when dipolar interaction is switched on. The main effect of dipolar interaction is to suppress the first order CB to fully-filled-lattice transition, dramatically shrinking the coexistence region between CB and SF or suppressing the CB-SF first-order transition altogether. For $ V_{\rm dip}=J$ (Fig.~\ref{FIG2}(a)), we observe that CB is surrounded by the SF phase. The CB-SF transition is of first order but, unlike the case of $ V_{\rm dip}=0$, the range of $\mu$ where we observe hysteretic behavior is small, corresponding to the thickness of the green solid line in Fig.~\ref{FIG2} (a). As a result, the coexistence region is highly suppressed. We also notice that, as expected, the CB solid is stabilized for smaller cavity-mediated interaction ($ V_{\rm ca}=2.5J$), compared to $ V_{\rm ca}=3.1J$ in the absence of dipolar interactions. 
As the dipolar interaction is increased to $ V_{\rm dip}=2J$ (Fig.~\ref{FIG2}(b)), the CB appears for $ V_{\rm ca} \sim 1.25$ J. Here, we observe a supersolid region separating the CB and SF. The CB-SF first order transition is therefore replaced by a two-step second order transition CB-CBSS and CBSS-SF.
In both phase diagrams of Fig.~\ref{FIG2}, we are unable to resolve the transition at fixed $n=0.5$ within the accuracy of our simulation results.

\section{Experimental realization}
Experiments with optical lattices within a high finesse cavity have been realized and have become popular in state-of-the-art laboratories. Quantum exotic phases from short-ranged contact interaction and infinite-range interaction mediated by the cavity has been observed using Rb \cite{EsslingerNature}. Such systems should be able to be extended to explore the quantum phases discussed above using magnetic polar atoms such as Cr, Er, and Dy \cite{chomaz_dipolar_2022}, and polar molecules, Er$_{2}$, KRb, NaK, NaRb \cite{bohn_cold_2017, christakis_probing_2022}, and Rydberg dressing \cite{schaus_observation_2012, hollerith_realizing_2022}.

The infinite-long range interactions can be tuned by adjusting laser intensity, while the dipolar interactions can be tuned by using different rotational states of the polar molecules, different projection state along the magnetic field quantization axes, and different Rydberg states for Rydberg dressing.
The filling factor and temperatures can be controlled by using different evaporation depths. By changing lattice light intensity, different tunnelings and chemical potentials can be realized. Time of flight and quantum gas microscopy can be used to detect the above discussed exotic quantum phases.

Here, we give a simple sketch of parameters used in experiments. For magnetic polar atoms Er and Dy in optical lattices using ultra-violet light $\sim$ 400 nm or less, the dipolar interaction can reach to 200 Hz. When tunneling rate is around 100 Hz, $V_{\textrm{dip}} / \textrm{J}$ is around 2. Thus, by tuning filling factor, SF, CB and CBSS are all able to be observed, when $V_{\textrm{ca}} / \textrm{J}$ is 2 and $U_{\rm s}/J$ = 20. When considering polar molecules, dipole moments can be as large as several Debye. This gives interactions of several kilo-Hertz to tens of kilo-Hertz even in lattice of 1064 nm laser light. However, low filling factors and high temperatures are two challenges. In this paper, though, we show that CBSS can be observed for filling as low as $n\sim 0.27$, which is  close to what is currently available. The temperatures needed to observe different phases and transitions are around one to several nano-Kelvins.

\label{sec:sec6}

\section{Conclusion}
\label{sec:sec7}
In conclusion, we have used quantum Monte Carlo by the worm algorithm to study the phase diagram of dipolar lattice bosons coupled to a high-finesse cavity in both hardcore and softcore case. The cavity-mediated infinite-range interaction enhances robustness of the checkerboard solid and supersolid. As a result, the checkerboard supersolid can already exist at filling factors as low as $\sim0.27$. The recent advent of molecular quantum gas microscope gives unprecedented resolution of the lattices \cite{christakis_probing_2022} which greatly facilitates the optimization of the cold polar molecule system. Therefore, we expect such a filling factor to be achieved in the near future, making the realization of the checkerboard supersolid phase with this setup possible. We also observed that cavity-mediated interactions facilitate observation of checkerboard solid and supersolid phases with magnetic atoms. As a consequence, based on the results presented, both polar molecules and magnetic atoms are good candidates to observe checkerboard supersolidity in optical lattices coupled to a high-finesse cavity within currently available experiments.

\begin{acknowledgments}
C. Zhang acknowledges support from the National Natural Science Foundation of China (NSFC) under Grant No. 12204173 and No. 12275263. The computing for this project was performed at the cluster at Clark University.
\end{acknowledgments}

\bibliography{cavity+dipolar}

\end{document}